\documentclass[11pt]{article}
\usepackage{moriond,graphicx}
\usepackage{pslatex}

\def\be{\begin{equation}}
\def\ee{\end{equation}}
\def\bea{\begin{eqnarray}}
\def\eea{\end{eqnarray}}

\def\mt{{m_t}}
\def\mts{{\mt^2}}
\def\mur{{\mu_r}}
\def\muf{{\mu_f}}
\def\mufs{{\muf^2}}
\def\shat{{\hat s}}
\def\GeV{\mbox{GeV}}
\def\msbar{\ensuremath{\overline{\mbox{MS}}}}
\def\mmu{{\overline{m}(\mur)}}
\def\z#1{\zeta(#1)}
\def\d#1{{d_{#1}}}
\def\eq#1{Eq.~\ref{#1}}
\def\fig#1{Fig.~\ref{#1}}
\def\tab#1{Tab.~\ref{#1}}
\def\LambdaQCD{\Lambda_{\mbox{\scriptsize QCD}}}
\begin{document}
\title{
\textnormal{\normalsize
            \phantom{m}\\
            HU-EP-10/23\quad DESY-10-066\hspace*{\fill}\hfill\\[2mm]
            }
\vspace*{40mm}
MEASURING THE RUNNING TOP-QUARK MASS}

\author{ ULRICH LANGENFELD }

\address{Humboldt-Universit\"at zu Berlin, Institut f\"ur Physik,
Newtonstra\ss e 15, D-12489-Berlin Germany }

\author{SVEN-OLAF MOCH}

\address{Deutsches Elektronen-Synchrotron, DESY, Platanenallee 6,
  D-15738 Zeuthen, Germany}

\author{PETER UWER \footnote{speaker at the conference}}

\address{Humboldt-Universit\"at zu Berlin, Institut f\"ur Physik,
Newtonstra\ss e 15, D-12489-Berlin Germany \includegraphics[scale=0.9]{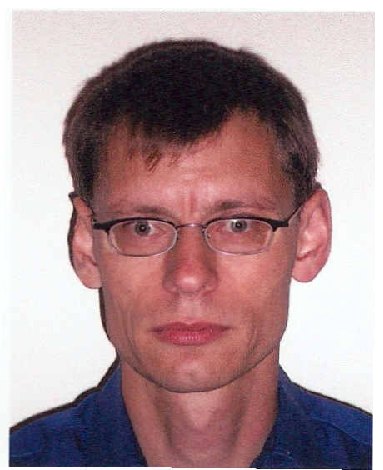}}

\maketitle\abstracts{
In this contribution we discuss conceptual issues of current mass
measurements performed at the Tevatron. In addition we propose an
alternative method which is theoretically much cleaner and to a large
extend free from the problems encountered in current measurements. In
detail we discuss the direct determination of the top-quark's running mass from
the cross section measurements performed at the Tevatron.}

\section{Introduction}
The top-quark is the heaviest known elementary particle discovered so far.  
It plays a prominent role in the physics program of the Tevatron 
accelerator at Fermilab and the Large Hadron Collider (LHC) at CERN 
(for recent reviews see e.g.~\cite{Bernreuther:2008ju,Incandela:2009pf}).  
The interest in top-quark physics stems from the fact that owing to its
large mass the top quark is a sensitive probe of the mechanism of
electroweak symmetry breaking. This is also the reason why the
top quark plays a special role in many extensions of the Standard
Model (SM)
aiming to give an alternative description of the mass generation. 
From the Standard Model viewpoint top-quark physics involves only
the mass and the matrix elements of the Cabibbo-Kobayashi-Maskawa
(CKM) matrix as free parameters in addition to the strong coupling
constant which we assume to be precisely measured by other means. 
Assuming that $V_{tb}$ is close to one---which is supported by
indirect measurements based on the assumption that only three flavour 
families exist---top-quark properties are thus precisely 
calculable in the SM provided the top-quark mass is known with good accuracy.
We also note that the large top-quark decay width $\Gamma_t\approx
1.5\GeV$ (a further consequence of the large mass) effectively cuts off
non-perturbative effects. As a consequence top-quark physics provides
an ideal laboratory for precise tests of the SM and its extension at
the scale of electroweak symmetry breaking.
The top-quark mass---a very fundamental property of the top quark---is
not only important for top-quark physics. It
enters as a very important parameter in electroweak fits constraining the
Standard Model, i.e. giving rise to indirect limits on the mass of the
Higgs boson (see e.g.~\cite{Flacher:2008zq}).
Currently, a value of $\mt = 173.1^{+1.3}_{-1.3} \GeV$ is quoted for
the mass of the top-quark~\cite{fnal:2009ec} (For an updated value
presented during the Moriond EW session
see~\cite{Lister,Peters}). 
This amounts to an experimental uncertainty of less than 1\%. 
Since the top-quark's width is so large that the top quark 
typically decays before it can hadronise~\cite{Bigi:1986jk} the
mass measurements proceed via kinematic reconstruction from the decay 
products and comparison to Monte Carlo simulations. However the
reconstruction of the four momentum of the coloured top quark from its 
uncoloured decay products introduces an intrinsic uncertainty due to
the non-perturbative mechanism of hadronisation in which the coloured
partons are transformed to colourless hadrons. There is a further
conceptual problem with the determination of the top-quark mass from
the kinematic reconstruction. Strictly speaking a higher-order
theoretical prediction of the observable under investigation is
required to extract a parameter of a model in a meaningful way. Only
beyond the Born approximation the renormalisation scheme can be
fixed. Thus, there is no immediate interpretation of the 
quantity currently measured at the Tevatron in terms 
of a parameter of the SM Lagrangian in a specific renormalization scheme.
A more detailed discussion will be given in section \ref{sec:topmass}.
In order to address this issue, we have chosen the following approach. 
We start from the total cross section for hadronic top-quark pair production, 
i.e. a quantity with well-defined renormalisation scheme dependence
which is known to 
sufficient accuracy in perturbative Quantum Chromodynamics (QCD).
Its dependence on the top-quark mass is commonly given in the on-shell scheme, 
although it is well-known that the concept of the pole mass has
an intrinsic theoretical limitation 
leading, for instance, to a poorly behaved perturbative series. 
This typically implies a strong dependence of the extracted value for the
top-quark mass on the order of perturbation theory. Similar effects
have been observed in $e^+e^-$ annihilation \cite{Hoang:2000yr}.
So-called short distance masses offer a solution to this problem. 
As we compute the total cross section as a function of the top-quark mass 
in the modified minimal subtraction (\msbar) 
scheme~\cite{Gray:1990yh,Chetyrkin:1999qi,Melnikov:2000qh} 
we demonstrate stability of the perturbative expansion and good properties of
apparent convergence~\cite{Langenfeld:2009wd}.
In particular, this allows for the direct determination of the 
top-quark's running mass from Tevatron measurements for the total 
cross section~\cite{Abazov:2009ae}, 
which is of importance for global analyses of electroweak precision data.
The direct extraction of the running mass also provides an important
cross check of the current measurements.
The outline of this contribution is as follows. In
section~\ref{sec:xsection} we briefly comment on the theoretical
status of the predictions for top-quark pair production. In section 
\ref{sec:topmass} we discuss in some details conceptual issues of
current measurements and how they can be avoided measuring the
top-quark mass in the \msbar\ scheme often called
the running mass for its dependence on the renormalisation scale.
The application is shown in section~\ref{sec:results}. A short summary
is given in section~\ref{sec:summary}.  

\section{The total cross section for top-quark pair production}
\label{sec:xsection}
We start by recalling the relevant formulae for the total cross section
$\sigma_{pp \to {t\bar t X}}$ of
top-quark hadro-production within perturbative QCD,
\begin{eqnarray}
  \label{eq:totalcrs}
  \sigma_{pp \to {t\bar t X}}(S,\mts) &=& 
  \sum\limits_{i,j = q,{\bar{q}},g} \,\,\,
  \int\limits_{4\mts}^{S}\,
  ds \,\, L_{ij}(s, S, \mufs)\,\,
  {\hat \sigma}_{ij}(s,\mts,\mufs)
  \, ,
\\
  \label{eq:partonlumi}
  L_{ij}(s,S,\mufs) &=& 
  {1\over S} \int\limits_s^S
  {d\shat\over \shat} 
  F_{i/p}\left({\shat \over S},\mufs\right) 
  F_{j/p}\left({s \over \shat},\mufs\right)
  \, ,
\end{eqnarray}
where $S$ denotes the hadronic center-of-mass energy squared and $\mt$ the
top-quark mass (taken to be the pole mass here). 
The standard definition for the parton luminosity $L_{ij}$ convolutes the two
parton distributions (PDFs) $F_{i/p}$ 
at the factorization scale $\muf$. Note that due to the additional
factor $1/S$ the fluxes at the Tevatron and the LHC can be directly
compared. The partonic cross sections 
${\hat \sigma}_{ij}$ 
parameterize the hard partonic scattering process after factorzation
of initial state singularities. Factoring out a common mass scale
squared $1/\mts$ the remaining part of the cross
section (often called scaling functions) 
depend only on dimensionless ratios of $\mt$, 
$\muf$ and the partonic center-of-mass energy squared $s$.

The QCD radiative corrections for the total cross section in 
\eq{eq:totalcrs}  
as an expansion in the strong coupling constant $\alpha_s$ 
are currently known completely at next-to-leading order 
(NLO)~\cite{Nason:1987xz,Beenakker:1989bq,Bernreuther:2004jv,Czakon:2008ii} 
and, as approximation, at next-to-next-to-leading order 
(NNLO)~\cite{Moch:2008qy,Moch:2008ai}.
The latter result is based on the known threshold corrections to the partonic
cross section ${\hat \sigma}_{ij}$, 
i.e. the complete tower of Sudakov logarithms in 
$\beta = \sqrt{1 - 4\mt^2/s}$ and the two-loop Coulomb
corrections, i.e. powers $1/\beta^k$ (see also~\cite{Beneke:2009ye} 
for some recent improvements).
It also includes the complete dependence on $\muf$ and the 
renormalization scale $\mur$, 
both being known from a renormalization group analysis.
The presently available perturbative corrections through 
NNLO lead to accurate predictions 
for the total hadronic cross section of top-quark pairs 
with a small associated theoretical 
uncertainty~\cite{Langenfeld:2009wd,Moch:2008qy,Moch:2008ai} 
(see also e.g.~\cite{Cacciari:2008zb} for related theory improvements 
through threshold resummation). For further refinements studied
recently we refer to 
\cite{Beneke:2009ye,Beneke:2009rj,Ahrens:2009uz,Ahrens:2010zv}.
We stress that aiming for a precision of the theoretical predictions
at the per cent
level also electroweak contributions need to be taken into account.
At the LHC these correction can amount up to 1--2\%, for details we
refer to~\cite{Beenakker:1993yr,Kuhn:2006vh,Bernreuther:2008md}. Very close
to the threshold the attractive part of the QCD potential may lead to
remnants of a would be boundstate~\cite{Hagiwara:2008df,Kiyo:2008bv}. 
These corrections affect
significantly differential distributions in the threshold region. A
prominent example is the $m_{tt}$-distribution, the invariant mass
distribution of the top-quark pair. Due to boundstate effects the
differential cross section obtains also a contribution from
kinematic regions below the nominal production threshold. 
If one could resolve this region
experimentally it would provide a sensitive method to measure the
top-quark mass similar to what is proposed for a future $e^+e^-$ linear 
collider.  The correction of the total cross section due to this
effect is of the order of 10 pb at the LHC. At the Tevatron where
colour octet production dominates this effect is less important.  

\section{The top-quark mass}
\label{sec:topmass}
We may start the discussion with a few general remarks.
When talking about the mass of an elementary
particle one should always keep in mind what is actually meant 
by this parameter.
This is in particular important for states which---due to
confinement---do not appear as
asymptotic states in the full field theoretical description. 
Since no free quarks exist we have to treat the quark mass similar to
any other parameter/coupling appearing in the underlying model.
In principle there is no difference between the treatment of the
coupling constant of the strong interaction $\alpha_s$ and the self
coupling of the quarks denoted by $\mt$. Note that we restrict our
selves to pure QCD and ignore the fact that the masses are generated
by the Higgs mechanism.
To measure a parameter of the Lagrangian we have to compare the
measurements with the theoretical predictions depending on the
unknown parameters of the theory. The theoretical prediction should be
as precise as possible so that a good agreement between data and
theory can be assumed provided the parameters are chosen (``fitted'') 
appropriate.
In particular one should use at least a next-to-leading order
prediction. There is a second even more important argument why at least a
next-to-leading order prescription is required: In leading-order no
precise definition of a parameter can be given. The difference between 
different definitions implemented by a specific renormalisation schemes 
is formally of higher order in perturbation
theory and thus only shows up when we go beyond the Born
approximation. To illustrate the point let us come back to
the quark mass. Two common schemes are frequently used in perturbation
theory. One is the on-shell or pole-mass scheme. The mass parameter in
the pole-mass scheme is defined as the location of the pole of the propagator. 
Since self-energy corrections
can shift the location the pole-mass
definition has to be enforced order by order in perturbation theory through the
renormalisation procedure. That is the renormalisation constants are
fixed order by order such that no shift in the renormalised pole mass occurs. 
Another scheme is the so-called modified minimal
subtraction scheme ($\overline{\mbox{MS}}$). This scheme is defined by 
subtracting the ultraviolet singularities appearing in the
unrenormalised theory order by order in a minimal way. That is just the
divergence itself (together with some irrelevant constants in case of
the modified MS) is absorbed into the redefinition of the bare
quantities. Since different renormalisation schemes
should be equivalent it must also be possible to convert from one scheme to
another. This is indeed the case. The relation between the pole mass $\mt$
and the $\overline{\mbox{MS}}$ mass $\mmu$ reads for example:
\begin{equation}
  \label{eq:massconversion}
  \mt = \mmu \* \left(1 + {\alpha_s(\mur)\over \pi} \d1 
    + \left({\alpha_s(\mur)\over \pi}\right)^2 \d2+\ldots\right).
\end{equation}
Treating $(n_f-1)$ flavours massless and expressing the QCD coupling
constant in the $n_f$-flavour theory through the coupling constant in
the $(n_f-1)$-flavour theory---that is using a scheme in which the running
of the coupling constant is solely determined by the massless
quarks---the constants $\d1, \d2$ read:
\begin{eqnarray} 
  \label{eq:d1def}
  \d1 &=& {4\over 3} + \ell, \\
  \label{eq:d2def}
  \d2 &=& {307\over 32} + 2 \* \z2 + {2\over 3} \* \z2 \* \ln 2 
  - {1\over 6}\*\z3 + {509\over 72}\*\ell + {47\over 24}\*\ell^2 
  \nonumber \\
  && 
  - \left( {71\over 144} + {1\over 3}\*\z2 + {13\over 36}\*\ell 
    + {1\over 12}\*\ell^2 \right)\*n_f 
  + {4\over 3}\sum_l\Delta(m_l/\mt)
  \, 
  ,
\end{eqnarray}
with $\ell=\ln\left({\mur\over \mmu}\right)$.
As mentioned before we observe in \eq{eq:massconversion} that the
difference between the pole mass and the running mass is formally
proportional to $\alpha_s$. 
We note that like $\alpha_s$ the $\overline{\mbox{MS}}$ mass depends on the
renormalisation scale. 
Since the top-quark mass is essentially measured at the Tevatron
from a kinematical fit the renormalisation scheme is not unambiguously
fixed. It is believed that the measured value should be interpreted as 
pole mass. 
However one should keep in mind that the reconstruction of the
top-quark momenta from the observed hadron momenta introduces a
further uncertainty due to colour reconnection which is expected to be 
of the order of 
$\LambdaQCD$. This is supported by a recent
study by Skands and Wicke where the influence of different models
for non-perturbative physics has been investigated~\cite{Skands:2007zg}. 
There is a further reason why the use of the pole mass should be
avoided when we are aiming for high accuracy. Qualitatively it is clear
that the full $\cal S$-matrix cannot have a pole at the location of
the quark mass since this would mean that the quark appears as
asymptotic state which is not the case due to confinement. A more
formal approach relates this uncertainty to a certain class of higher 
order corrections spoiling the convergence of the 
perturbative series
\cite{Bigi:1994em,Beneke:1994sw}.
Technically the problem becomes manifest when one uses a Borel
summation of the perturbative series. The back transformation of the
Borel transform is 
ill-defined due to the existence of a pole on the real axis. 
Taking the residue of the pole as an estimate of
the theoretical uncertainty it is found that an ambiguity of the order
of $\LambdaQCD$ is introduced. That is, the pole mass scheme has an intrinsic
uncertainty of the order of $\LambdaQCD$ \cite{Beneke:1994sw}: 
It is thus conceptually
impossible to measure the pole mass with an accuracy better than $\LambdaQCD$.

Taken the last statements into account, a theoretical clean approach
to measure the
top-quark mass is to choose a specific observable, calculate the higher
order corrections choosing a well defined renormalisation scheme like
for example the running mass and then to compare with the
measurements. This idea has been pursued in  \cite{Langenfeld:2009wd}.
As observable the inclusive cross section has been used.
In the next section we will comment on the details of this approach.

\section{The cross section using the \msbar\ mass}
\label{sec:results}
\begin{figure}[htbp]
  \begin{center}
    \includegraphics[width=0.47\textwidth]{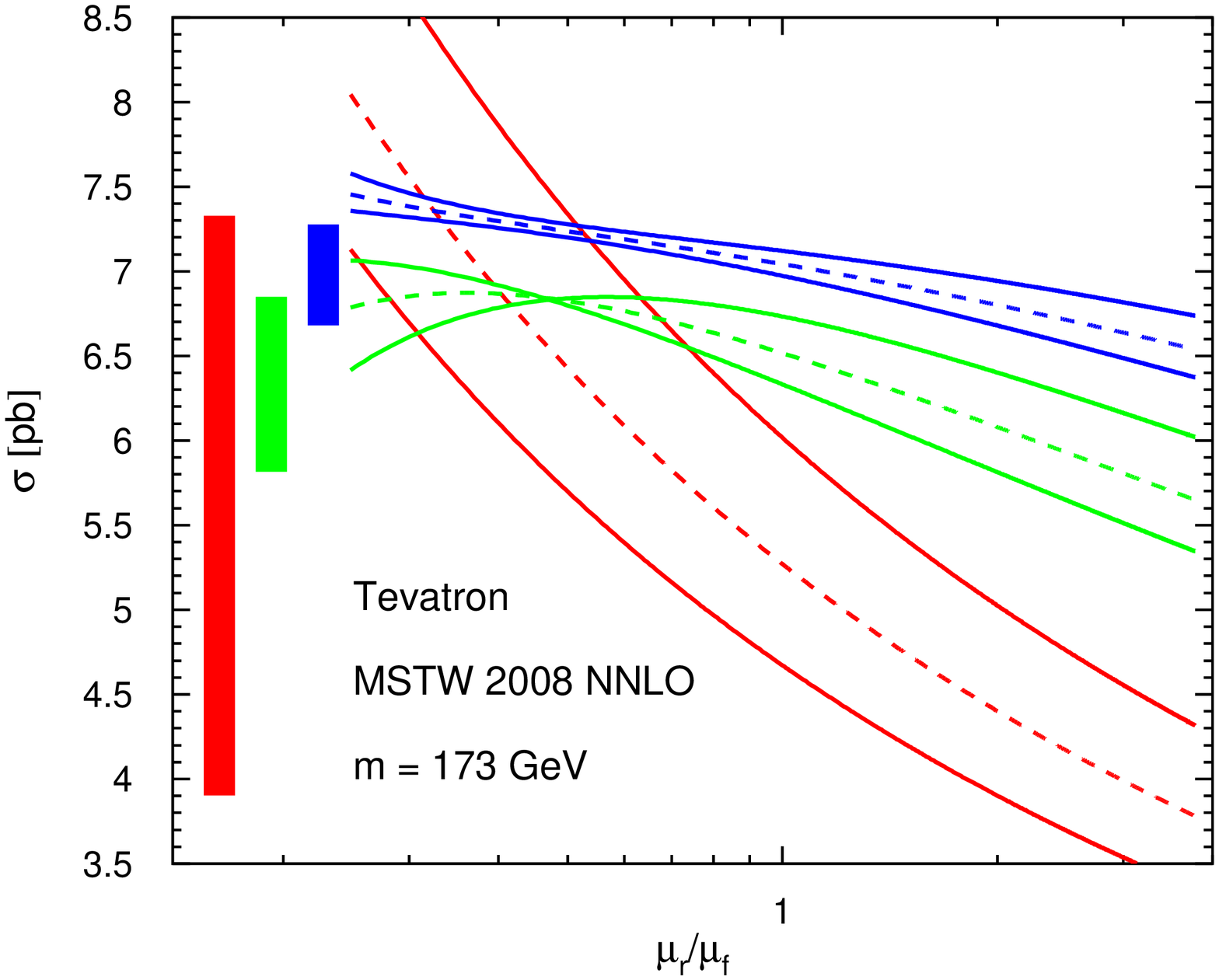}
    \hspace*{0.3cm}\includegraphics[width=0.47\textwidth]{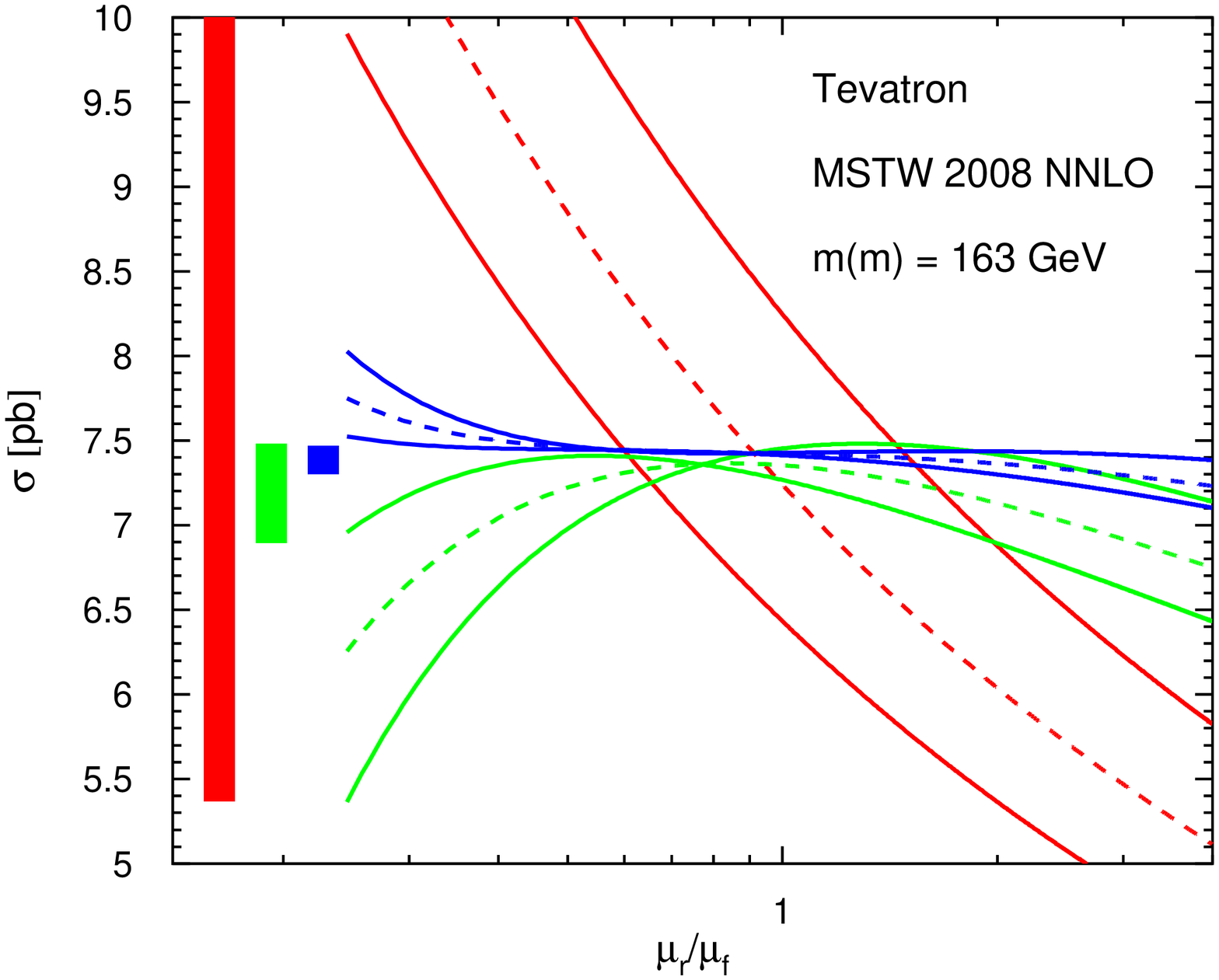}
    \caption{Cross section predictions using the pole mass (left) and
    the \msbar\ mass right as function of the renormalisation scale
    for three different factorisations scales $\muf=0.5 m, m, 2m$.}
    \label{fig:xsectionsPolevsMS}
  \end{center}
\end{figure}
\begin{figure}[htbp]
  \begin{center}
    \includegraphics[width=0.7\textwidth]{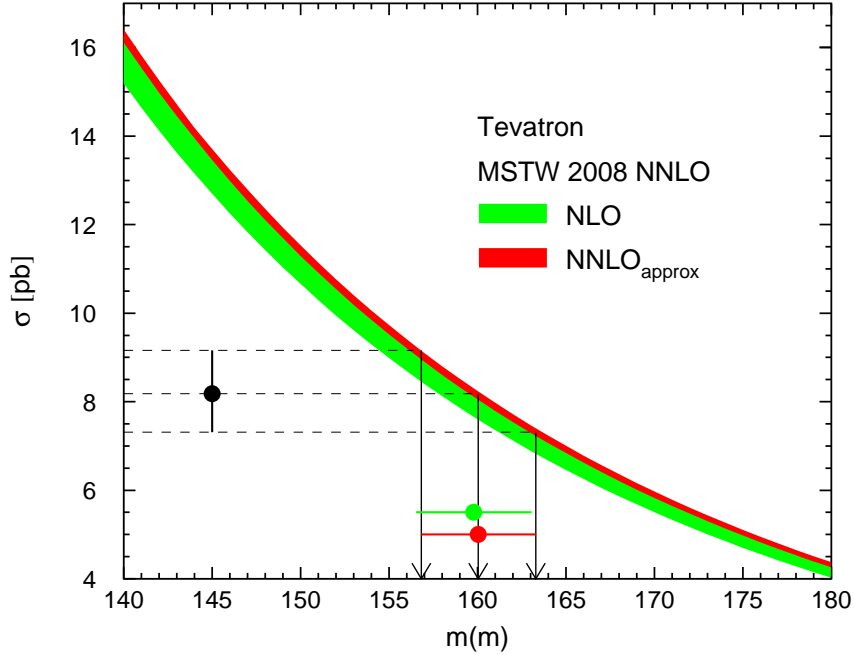}
    \caption{Cross section predictions using the \msbar\ mass as 
      function of the top quark mass.
      \label{fig:MassDetermination}
      }
  \end{center}
\end{figure}

As outlined in the previous section the main idea to circumvent the
aforementioned problems of the current experimental determination of
the top quark mass is to choose a sensitive observable translated to
the \msbar\ scheme as far as the mass parameter is concerned. The mass
value is than obtained from a direct comparison with experimental
data. In \cite{Langenfeld:2009wd} the results for the total cross
section \cite{Moch:2008qy} were translated to the \msbar\ scheme using
\eq{eq:massconversion} and \eq{eq:d1def}. The translation is first
done at a fixed renormalisation scale for three different
factorisation scales. The full renormalisation scale dependence is
recovered from a renormalisation group analysis.
In \fig{fig:xsectionsPolevsMS} the cross section is shown for three
different choices of the factorisation scales $\muf=0.5m,m,2m$ as
function of the renormalisation scale $\mur$. The left plot shows the
cross section using a pole mass of 173~GeV. The right plot employs the
running mass definition with a mass value $m(m)=163$~GeV. The bands at the
left side of the two plots show an estimate of what one may call a theoretical
uncertainty. They are obtained by varying the relative
scales $\mur/m$ and $\muf/m$ between 0.5 and 2. We note that 
there is typically 
a crossing of the different curves for a given order. In particular the
central scale is not necessarily between the two extreme scales. This
behaviour appears when the central scale corresponds to a
plateau. If one studies the uncertainty bands two important features can
be observed. Compared to the pole mass scheme the cross section 
prediction using  the \msbar\ mass is much more stable. The NLO band overlaps
with the  NNLO band, in fact the NNLO band is fully included in the NLO
band. Furthermore the size of the bands is reduced compared to the
predictions using the pole mass.    
The perturbative prediction becomes thus much more stable with respect to
radiative corrections.
Using the cross section to determine the mass parameter this leads to
a much more stable determination in the running mass scheme compared
to a determination in the pole mass scheme. 
In \fig{fig:MassDetermination} the cross section is shown as a
function of the \msbar\ mass evaluated at $\mur=m$. The wide band is
the NLO prediction while the narrow band is an approximation to the
full NNLO result. The uncertainty bands are again due to a variation of the
scales. The data point shown to the left is the recent Tevatron 
measurement \cite{Abazov:2009ae} for the cross section:
\begin{equation}
  \label{eq:TevatronXSection}
  \sigma = 8.18^{+0.98}_{-0.87} \mbox{ pb}.
\end{equation}
We note that
this measurement effectively depends on an assumed top-quark mass since
detector efficiencies and other systematics are estimated from Monte
Carlo simulations using
a specific mass. In principle this dependence is known and can be
taken into account. The dependence is however rather mild and thus
does not give a significant shift in the cross section. In the current analysis
it is not taken into account.
The extraction of the top-quark mass in the \msbar\ mass is now 
straightforward. Projecting the measured value on the curves we can
immediately read off the corresponding mass value. An illustration of
this procedure is visualized in \fig{fig:MassDetermination}.
The outcome of this procedure is presented in \tab{tab:massnumbers}.
For comparison we also show the results for the case that the pole
mass is used. We observe that the extraction in the \msbar\ scheme
leads---as anticipated already---to very stable results with respect 
to different orders of the
perturbative prediction. The determination using the pole mass scheme
however shows large differences when going from LO to NLO and finally
to NNLO.
\renewcommand{\arraystretch}{1.35}
\begin{table}[ht!]
\begin{center}
\caption{ \label{tab:massnumbers}
  The  LO, NLO and approximate NNLO results for the top-quark mass in
  the \msbar\ scheme ($m(m)$) and the pole mass scheme ($\mt$) for the 
  cross section measured at Tevatron.%
  }
\begin{tabular}{|l|c|c|}
  \hline
  &  $m(m)$ [$\GeV/c^2$] 
  &  $\mt$ [$\GeV/c^2$]
  \\ \hline
  LO 
  &$159.2^{+3.5}_{-3.4}$
  &$159.2^{+3.5}_{-3.4}$
  \\
  NLO
  &$159.8^{+3.3}_{-3.3}$
  &$165.8^{+3.5}_{-3.5}$
  \\
  NNLO
  &$160.0^{+3.3}_{-3.2}$
  &$168.2^{+3.6}_{-3.5}$
  \\
  \hline 
\end{tabular}
\end{center}
\end{table}
As final result the value corresponding to the NNLO approximation is
quoted:
\begin{equation}
  \label{eq:result}
  m(m) = 160^{+3.3}_{-3.2}~\GeV/c^2.   
\end{equation}
Converting the
running mass to the on-shell mass yields a result which is consistent with the
direct measurements at Tevatron. 
Due to the weak sensitivity of the cross section
with respect to the mass the method is not competitive with the direct
measurements as far as the uncertainty is concerned, however the
method provides an independent cross check and is theoretically rather clean.  

\section{Summary}
\label{sec:summary}
The current top-quark mass measurements at the Tevatron claiming an
accuracy at the per cent level suffer from various uncertainties
(for a similar discussion see also \cite{Smith:1996xz}):
\begin{enumerate}
\item The renormalisation scheme is not uniquely defined since the
  measurement is based on a kinematic reconstruction without relying
  on higher-order predictions required to define unambiguously a
  specific renormalisation scheme.
\item The kinematic reconstruction of the top-quark momentum from the
  momenta of the decay products introduces an additional uncertainty
  due to the non-perturbative aspects of colour reconnection. The
  naive estimate that the uncertainty is of the order of $\LambdaQCD$
  is supported by phenomenological studies \cite{Skands:2007zg} where
  the uncertainty was estimated to be of the order of 500 MeV.
\item The pole mass itself has an intrinsic uncertainty of the order of 
  $\LambdaQCD$ which is usually attributed to IR renormalons. 
\end{enumerate}
One should note that each of the problems itself is hard to improve if
not impossible. The intrinsic uncertainty of the pole mass for example 
cannot be improved. 
As a consequence we advocate an alternative method to determine the 
top-quark mass which is to a large extend free from the aforementined problems.
The basic idea is to extract the mass---as it is done in general for any
parameter in a theoretical model---from a detailed comparison of the value
of an experimentally measured observable with the theoretical
predictions therefore. This leads to a clean definition of the renormalisation
scheme adopted for the mass parameter. Using in addition a
short distance mass like the \msbar\ mass  the intrinsic uncertainties
of the pole mass are circumvented.
Along these lines we have used the total cross section written in
terms of the \msbar\ mass to extract the top-quark mass from the cross
section measurements at Tevatron. 
Our final result for the top-quark mass $m(m)$ in the \msbar\ scheme 
derived from the cross section measurements at the Tevatron is
presented \eq{eq:result}.
We find a remarkable stability with respect to the perturbative order
of the theoretical predictions.
Converted to the pole mass scheme the value is consistent with direct
measurements. However we stress that despite the large uncertainty due
the poor sensitivity of the total cross section with respect to the
mass the result is theoretically rather clean and in particular free
of uncertainties which are not quantified in the direct measurements.

\section*{Acknowledgments} P.U. would like to thank the organisers for the
  invitation to present this work at the Moriond EW conference 2010.

\end{document}